\documentclass[11pt]{article}

\usepackage{graphicx}
\usepackage{citesort}
\usepackage{amssymb}
\usepackage{amsfonts}
\usepackage{amsmath}
\usepackage{epsfig}
\usepackage{fancybox}
\usepackage{textcomp}
\usepackage{multirow}
\usepackage{appendix}
\usepackage{setspace}

\usepackage[top=1.5cm,bottom=1.5cm,left=1.5cm,right=1.5cm]{geometry}
\renewcommand{\baselinestretch}{1.65}
    \def\Complex{{\rm\rule[.23ex]{.03em}{1.1ex}\kern-.3em{C}}}

    \newcommand{\be}{\begin{equation}} \newcommand{\ee}{\end{equation}}
    \newcommand{\bea}{\begin{eqnarray}} \newcommand{\eea}{\end{eqnarray}}
    \newcommand{\benum}{\begin{enumerate}} \newcommand{\eenum}{\end{enumerate}}
    \def\proof{\noindent\hspace{2em}{\itshape Proof: }}
    
    \def\endproof{\hspace*{\fill}~$\square$\par\endtrivlist\unskip}

  \newtheorem{theorem}{Theorem}
  \newtheorem{definition}{Definition}
  
  \newtheorem{lemma}{Lemma}
  \newtheorem{proposition}{Proposition}

\newtheorem{Corollary}{Corollary}

\begin{document}

\title{\LARGE\bf Self-Optimized OFDMA\\via Multiple Stackelberg Leader Equilibrium\thanks{This work was supported in part by EPSRC under grant EP/E022308/1.}}

\setcounter{footnote}{1}

\author{Jie Ren\thanks{School of Electronics and Information Engineering, Beijing Jiaotong University, China.}, Kai-Kit Wong\thanks{Department of Electronic and Electrical Engineering, University College London, London, WC1E 7JE, United Kingdom.}, and Jianjun Hou$^\ddag$}

\maketitle
\thispagestyle{empty}

\begin{abstract}
The challenge of self-optimization for orthogonal frequency-division multiple-access (OFDMA) interference channels is that users inherently compete harmfully and simultaneous water-filling (WF) would lead to a Pareto-inefficient equilibrium. To overcome this, we first introduce the role of environmental interference derivative in the WF optimization of the interactive OFDMA game and then study the environmental interference derivative properties of Stackelberg equilibrium (SE). Such properties provide important insights to devise free OFDMA games for achieving various SEs, realizable by simultaneous WF regulated by specifically chosen operational interference derivatives. We also present a definition of all-Stackelberg-leader equilibrium (ASE) where users are all foresighted to each other, albeit each with only {\em local} channel state information (CSI), and can thus most effectively reconcile their competition to maximize the user rates. We show that under certain environmental conditions, the free games are both unique and optimal. Simulation results reveal that our {\em distributed} ASE game achieves the performance very close to the near-optimal {\em centralized} iterative spectrum balancing (ISB) method in \cite{Yu-06}.

\begin{center}
{\bf Index Terms}
\end{center}
Game theory, Nash equilibrium, OFDMA, Self-optimization, Stackelberg equilibrium.
\end{abstract}

\thanks{\singlespacing}
\setcounter{page}{0}

\newpage
\section*{\sc I. Introduction}
Dynamic spectrum access enabled by cognitive radio technologies has recently attracted much attention \cite{Mitola-99,Haykin-05,Mishra-06}. Of particular interest is the orthogonal frequency-division multiple-access (OFDMA) interference channel model in which multiple communication links coexist and share a spectrum of frequency bands dynamically without a centralized spectrum manager.\footnote{For infrastructure-based communications where a centralized spectrum manager knowing {\em full} channel state information (CSI) between {\em every} transmitter-receiver link is present, optimal approaches are well known \cite{Cendrillon-06,Yu-06,Han-05}, but they are hardly practical for infrastructureless OFDMA systems if a large number of users are involved, let alone its susceptibility to CSI errors.} The model is motivated by its suitability to analyze multicell networks and the move of regulatory bodies in favor of liberalization and license-exempt spectrum policies for higher utilization. In interference channels, often infrastructureless, users are all {\em uncoordinated} individuals, and inherently compete with each other. How individuals gather sufficient network information and optimize themselves to benefit not only their own but the entire network is an open challenge.

In this regard, game theory has emerged as a unique methodology to analyze a group of self-organized mutually competing users and can be adopted to design autonomous access control methods for OFDMA, e.g., \cite{Yu-02,Etkin-07,Lee-06,Lai-08,Peyman-2009,Scutari-08}. However, as reviewed in \cite{Kim-11}, there are fundamental deadlocks that prevent existing results from making substantial impacts. The first deadlock is the low efficiency of Nash equilibrium (NE) where users are {\em myopic} and harmfully compete with each other \cite{Popescu-07}. NE is also not unique and its performance is unpredictable. To avoid over-competition, Stackelberg equilibrium (SE) arises where {\em foresighted} users or {\em leaders} can benefit more from the game, by knowing full CSI and the strategies of all other myopic users or {\em followers}. In \cite{Su-09TWC}, one-leader SE (OSE) for OFDMA was studied revealing a large gain over NE.

Unfortunately, there are operational obstacles in achieving OSE. First, to be qualified as a leader, the user should possess tremendous cognition capability, which, albeit, could be learned from the environment by conjectures \cite{Su-09TSP}.\footnote{Note that for OSE, while conjectural equilibrium (CE) can be used to make the leader learn the necessary network-wide information, but it was acknowledged in \cite{Su-09TSP} that extending it to $K$-SE scenarios is a topic for future investigation.} Worse, obtaining the OSE would require playing a bi-level game,\footnote{To achieve OSE via a bi-level game, the leader will start his action first, wait for all the followers to play a sub-game to reach an NE, then revise his action, and the whole process repeats until convergence. A bi-level game is needed even for CE.} in which there is a strict order of how the game proceeds to reach the designated equilibrium. For OFDMA of uncoordinated users, this is unrealizable. Another long-standing issue is that OSE is not known to be unique.

The challenges do not end there, as there is strong desire to extend the SE framework to the cases where multiple leaders coexist \cite{Su-09TSP,Larsson-09,Xu-09,Kim-2010}. In \cite{Xu-09}, DeMiguel {\em et al.}~studied the extension and in their definition of $K$-SE ($K$ leaders out of all users), leaders are myopic with respect to other leaders, but foresighted to followers. To approach the $K$-SE, usual challenges in OSE remain. Further, according to their definition, if all users are leaders, the game degenerates to an NE, losing the foresightedness advantages of the leaders, making it inefficient for OFDMA systems. Most recently, $K$-SE was employed for femtocell communications in \cite{Kim-2010} where strong assumptions on leaders such as full CSI\footnote{In this paper, {\em full} CSI refers to the CSI knowledge of every transmitter-receiver link of the entire network, in contrast to {\em local} CSI which we refer to it as the CSI only for a transmitter to its intended receiver.} and exhaustive optimization, apply.

The paper aims to achieve a truly {\em self-optimized} OFDMA network where every transmitter individually optimizes its own power and subcarrier allocation, based on its {\em local} CSI and interference observations, for maximizing the network rates. To this end, we first revisit the water-filling (WF) optimization of a user's parameters by introducing the role of {\em interference derivative} which measures the environmental changes with respect to a change in the power allocation strategy of that user. We show that the environmental interference derivative plays a crucial role in the uniqueness of the OFDMA game, and by analyzing the environmental interference derivative property of NE and OSE, we can construct a {\em free} OSE game that can be approached by simultaneous (or iterative) WF\footnote{It is worth pointing out that in our proposed scheme, the WF solution incorporates the operational interference derivative into it and is fundamentally different from the conventional WF and the modified WF in \cite{Yu-07itaw}.} in an arbitrary order, regulated by an operational interference derivative. As a final outcome, we present a new all-leader SE (ASE) definition under which users are all foresighted to each other with their belief that others are myopic. By extending the analysis, we devise a free ASE game that is achievable again by a regulated simultaneous WF interaction. We also show that under certain environmental conditions, the proposed free games have unique equilibriums and optimal. In the context of OFDMA, this paper has made the following major contributions:
\begin{itemize}
\item We provide a new sufficient condition for the uniqueness of NE.
\item We prove that OSE is unique and a free OSE game can be constructed.
\item We define a new ASE and propose a free ASE game which is unique under certain conditions.
\item The free games can be achieved by simultaneous generalized WF (GWF) empowered by the specifically chosen operational interference derivatives, with the aid of only users' local CSI.
\item Simulation results reveal that the proposed free ASE game achieves the sum-rate very close to that of the {\em centralized} iterative spectrum balancing (ISB) method for the OFDMA channel in \cite{Yu-06}.
\item Our analysis provides the sufficient conditions for convergence to NE, OSE and ASE.
\end{itemize}

\section*{\sc II. Preliminaries}
\subsection*{A. OFDMA Interference Channels}
\begin{figure}[]
\begin{center}
\includegraphics[width=8cm]{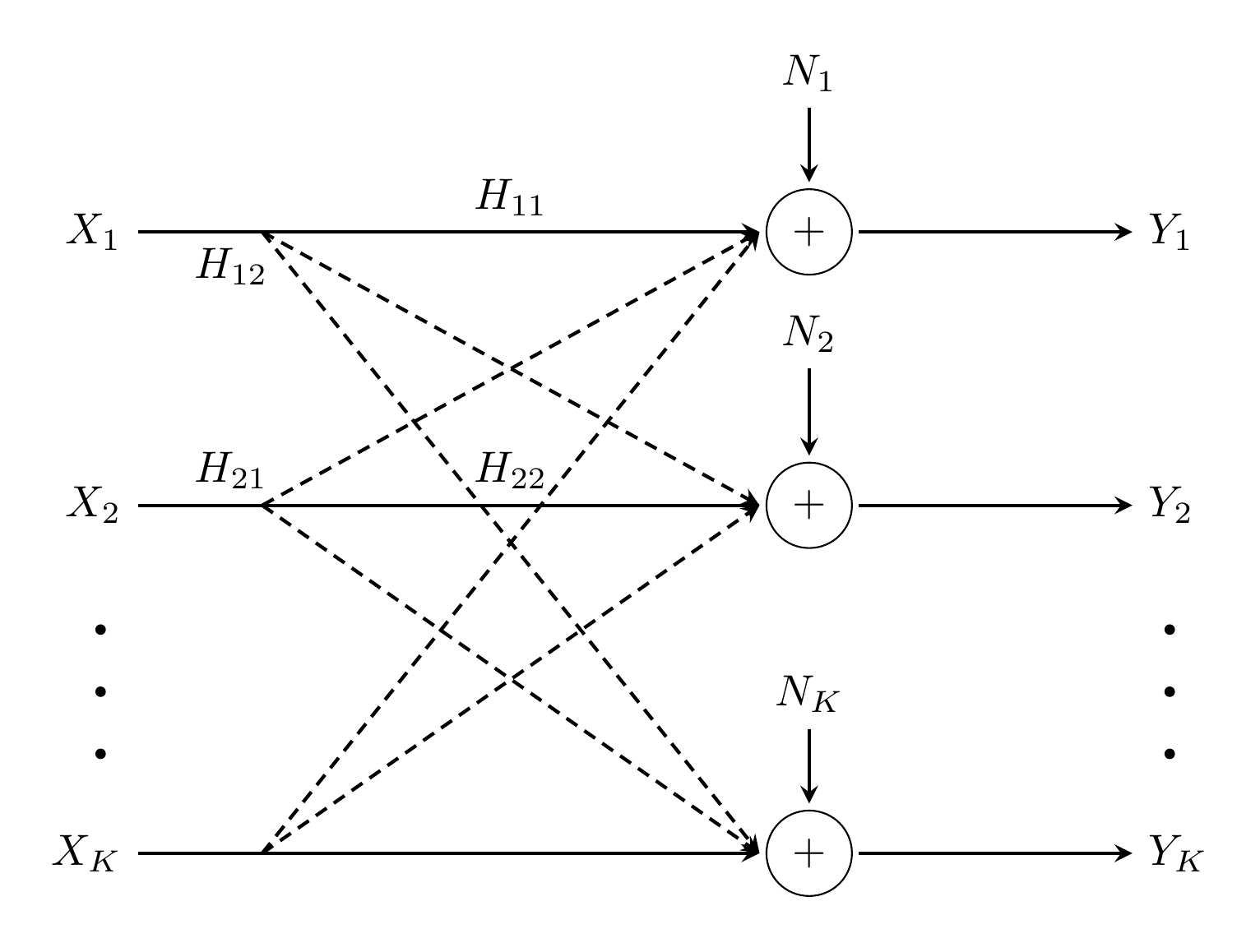}
\caption{A multi-user interference channel model.}\label{fig_gicm}
\end{center}
\end{figure}

Consider a $K$-user OFDMA infrastructureless system, as shown in Fig.~\ref{fig_gicm}, where each user is free to occupy any of the $N$ orthogonal subcarriers for communications. The users operate in a non-cooperative manner and inherently compete with each other. For user $k$, the total transmitted power is constrained by
\begin{equation}\label{pmax}
\sum_{n=1}^Np_k[n]\le P_k,~\forall k \in \{1,2,\dots,K\},
\end{equation}
where $p_k[n]$ denotes the power allocated for the $n$th subchannel by user $k$ and $P_k$ denotes the maximum total power for user $k$. We write ${\sf p}_k\triangleq \{p_k[1],p_k[2],\dots,p_k[N]\}$ as the power allocation pattern of user $k$, which is drawn from some power allocation strategy ${\cal P}_k$, or denoted by ${\sf p}_k\in{\cal P}_k$.

Let $H_{ij}[n]$ denote the flat-fading channel coefficient from transmitter $i$ to receiver $j$ and $N_k[n]$ denote the noise power density for the complex additive white Gaussian noise (AWGN) at receiver $k$ on the $n$th subchannel. The achievable rate for user $k$ can therefore be given by
\begin{equation}\label{eqn:Rk-2}
R_k\equiv\sum_{n = 1}^NR_k[n]
=\sum_{n = 1}^N {\log _2\left(1 + \frac{{p_k[n]}}{{\sigma _k[n] + I_k[n] }}\right)}=\sum_{n = 1}^N {\log _2\left(1 + \frac{{p_k[n]}}{{c_k[n]}}\right)},
\end{equation}
where $\sigma_k[n]\triangleq \frac{{N_k[n] }}{{\left| {H_{kk}[n] } \right|^2 }}$ is the normalized noise power on subchannel $n$, $I_k[n]=\sum_{i = 1\atop i \ne k}^K p_i[n]\theta_{ik}[n]$ is the total interference power on subchannel $n$ for user $k$, with $\theta_{ik}[n]\triangleq
\frac{|H_{ik}[n]|^2}{|H_{kk}[n]|^2}$ denoting the normalized (by user $k$) channel power gain from transmitter (or interference) $i$ to receiver $k$, and $c_k[n]\triangleq\sigma_k[n]+I_k[n]$. For transmitter $k$, it possesses only the knowledge of local CSI which corresponds to $\{c_k[n]\}$ for all $n$.

\subsection*{B. Input-Output Subsystem: Environmental Interference Derivative}
In this setup, users are all uncoordinated individuals and each user will allocate its power over the subcarriers to maximize its own rate based on its observation of the environment and its belief on how the environment would react to its action. The environment user $k$ observes can change because other users may alter their strategies to respond to the environmental changes caused by its own action. As such, it is a dynamic process where users all interact and could converge to a compromised equilibrium such as NE.

\begin{figure}[]
\begin{center}
\includegraphics[width=8cm]{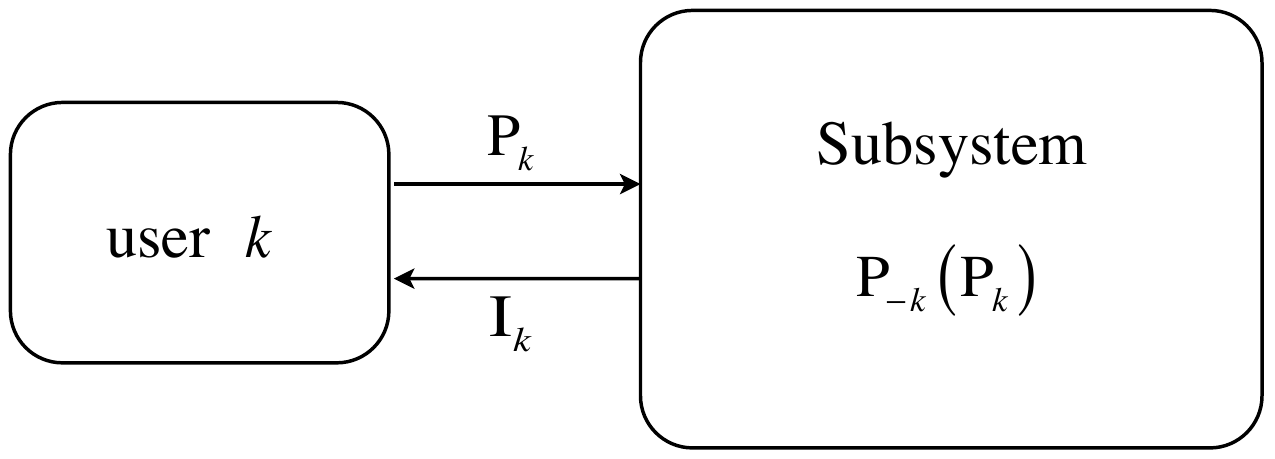}
\caption{The power game subsystem.}\label{fig_pgs}
\end{center}
\end{figure}

To model this interaction, from user $k$'s viewpoint, we can regard other users as a subsystem which takes its power allocation at time $t$ as inputs (i.e., $\{p_k^t[n]\}$) and produces a new interference pattern at time $t+1$ as outputs (i.e., $\{I_k^{t+1}[n]\}$), as shown in Fig.~\ref{fig_pgs}. Under this model, we define the {\em environmental interference derivative}, $\varphi_k^t[n]\triangleq\frac{\partial I^{t+1}_k[n]}{\partial p_k^t[n]}$, which measures the environmental change on a subcarrier seen by a user, caused by a change in the user's own power allocation on that subcarrier. This captures the essence of the overall network response and is a key parameter that determines whether or not such user interaction converges, and helps develop self-optimized algorithms for achieving desirable equilibriums (e.g., ASE).

\subsection*{C. GWF and Operational Interference Derivative}
Let $\boldsymbol{\mathcal{P}}=\{{\cal P}_1,{\cal P}_2,\dots,{\cal P}_K\}$ be the set of all users' power allocation strategies and ${\sf p}_k\in{\cal P}_k~\forall k$. We define ${\sf p}_{-k}\triangleq \{{\sf p}_1,\dots,{\sf p}_{k-1},{\sf p}_{k+1},\dots,{\sf p}_K\}$ embracing the power allocation of all users except user $k$. User $k$'s power allocation can be optimized by solving the following rate maximization problem (with $t$ omitted):
\begin{equation}\label{eqn:BP_k}
{\sf p}_k^*=\arg\max_{{\sf p}_k\in\mathcal{P}_k}\sum_{n=1}^N\log_2\left(1+\frac{p_k[n]}{c_k[n]}\right)~~\mbox{s.t.}~~\sum_{n=1}^Np_k[n]\le P_k,
\end{equation}
which can be solved by a Lagrangian multiplier formulation
\begin{equation}
\mathcal{L}=\sum_{n=1}^N\ln\left(1+\frac{p_k[n]}{c_k[n]}\right)-\lambda\left(\sum_{n=1}^Np_k[n]-P_k\right)
\end{equation}
with the Lagrange multiplier $\lambda$. To proceed, it can be easily shown that\footnote{The derivation can be done by recognizing that $\frac{\varphi+1}{c+p}-\frac{\varphi}{c}=\frac{1}{\frac{(c+p)c}{c-\varphi p}}=\frac{1}{\frac{(c+p)c}{c-\varphi p}-p+p}=\frac{1}{\frac{c^2+\varphi p^2}{c-\varphi p}+p}$.}
\begin{equation}
\frac{\partial{\cal L}}{\partial p_k[n]}=\frac{\varphi_k[n]+1}{c_k[n]+p_k[n]}-\frac{\varphi_k[n]}{c_k[n]}-\lambda=\frac{1}{\eta_k[n]+p_k[n]}-\lambda,
\end{equation}
where
\begin{equation}\label{eqn:etak}
\eta_k[n]\triangleq\frac{c_k^2[n]+\varphi_k[n]p_k^2[n]}{c_k[n]-\varphi_k[n]p_k[n]}
\end{equation}
is the ``interactive'' network noise due to $p_k[n]$ through the environmental interference derivative.

The Karush-Kuhn-Tucker (KKT) conditions for optimality suggest a WF strategy
\begin{equation}\label{gwfl}
p_k[n]=(w_k-\eta_k[n])^+,
\end{equation}
where ${(a)}^+=\max(0,a)$ and $w_k=\frac{c_k^2[n]+\varphi_k[n]p_k^2[n]}{c_k[n]-\varphi_k[n]p_k[n]}+p_k[n]$, for $p_k[n]>0$, is chosen to satisfy the user's power constraint $\sum_np_k[n]\le P_k$, and interpreted as the ``water-level'' of the solution. Incorporating the interactive nature of the system, the optimal power allocation for user $k$ can be realized by
\begin{equation}\label{eqn:pkWF}
p_k^{t}[n]=\left(w_k^t-\frac{\left(c_k^t[n]\right)^2+\varphi^t_k[n]\left(p_k^{t-1}[n]\right)^2}{c_k^t[n]-\varphi_k^t[n]p_k^{t-1}[n]}\right)^+.
\end{equation}
As a result, to use the above solution (\ref{eqn:pkWF}), the user needs to have a belief on how the environment would react, an {\em operational} interference derivative, or a belief on $\varphi^t_k[n]$, denoted as $\tilde{\varphi}_k[n]$. For instance, NE is defined such that at the equilibrium (with the corresponding parameters marked with $^*$) it satisfies
\begin{equation}\label{eqn:NE}
R_k({\sf p}_k^*,{\sf p}_{-k}^*)\ge R_k({\sf p}_k,{\sf p}_{-k}^*)~\forall k.
\end{equation}
Since the rate maximization for an NE user is performed with respect to some fixed interference pattern, $\{I_k({\sf p}_{-k}^*)[n]\}$, each user has a belief that the environment will remain static following a change of its power allocation strategy. In other words, $\tilde{\varphi}_k[n]=0$, leading to the well-known simultaneous WF \cite{Popescu-07}
\begin{equation}\label{eqn:pkWF-ne}
p_k^{t}[n]=\left(w_k^t-c_k^t[n]\right)^+~\forall k.
\end{equation}

The operational interference derivative is the key to which user's foresightedness can be nurtured. By a proper choice of the operational interference derivative it is possible to create a free OFDMA game for approaching a more desirable equilibrium, which we will address next. It is worth emphasizing that the GWF in (\ref{gwfl}) is fundamentally different from the modified WF in \cite{Yu-07itaw}. The main difference is that in our GWF there is an {\em interactiveness} through $\eta_k[n]$ which does not exist in the WF in \cite{Yu-07itaw}.

\section*{\sc III. SE and Its Free Game Realization}
\subsection*{A. New Key Results of NE}
Depending on how users react to the environmental changes, the system may converge to an equilibrium. Throughout, we use the superscript $*$ to denote the parameters at the equilibrium so the overall strategy is ${\bf P}^*=\{{\sf p}_k^*,{\sf p}_{-k}^*\}$. NE systems are Pareto-inefficient \cite{Yu-02,Etkin-07,Lee-06,Lai-08,Popescu-07}, which has motivated the concept of SE in which there can be foresighted users who can maximize their rates more effectively by knowing other myopic users' actions. In light of this, this paper aims to devise self-optimization methods that can lead the OFDMA game to an SE. To do so, we first develop some new results of NE and consider the scenario where user $\kappa$ is the user of interest and other users, for $i\ne\kappa$, are at NE, with user $i$'s power allocation denoted by ${\sf NE}_i({\sf p}_\kappa)=\{NE_i({\sf p}_\kappa)[1],\dots,NE_i({\sf p}_\kappa)[N]\}$, which is a function of user $\kappa$'s power allocation.

\begin{lemma}
The rate of change of the power allocation for user $i$ with respect to that for user $\kappa$ is given by
\begin{equation}\label{eqn:varpi-ne1}
\frac{\partial NE_i({\sf p}_\kappa)[n]{\sf sgn}(p_i[n])}{\partial p_\kappa[n]}=-\theta_{\kappa i}{\sf sgn}(p_i[n]),
\end{equation}
where ${\sf sgn}(x)$ returns one if $x>0$ or zero if $x\le 0$.
\end{lemma}

\proof First, assume that $p_i[n]>0$. Then, by chain rule, we have
\begin{equation}
\frac{\partial NE_i({\sf p}_\kappa)[n]}{\partial p_\kappa[n]}=\frac{\partial NE_i({\sf p}_\kappa)[n]}{\partial I_i[n]}\frac{\partial I_i({\sf p}_\kappa)[n]}{\partial p_\kappa[n]}.
\end{equation}
As user $i$ is using an NE strategy, we have, from (\ref{eqn:pkWF-ne}), that $p_i[n]=NE_i({\sf p}_\kappa)[n]=w_i-\sigma_i[n]-I_i[n]>0$ and therefore, $\frac{\partial NE_i({\sf p}_\kappa)[n]}{\partial I_i[n]}=-1$. On the other hand, by definition, we have $\frac{\partial I_i[n]}{\partial p_\kappa[n]}=\theta_{\kappa i}[n]$. Finally, note that if $p_i[n]=0$, the rate of change will be zero. Therefore, (\ref{eqn:varpi-ne1}) is obtained, which completes the proof.\endproof

\begin{lemma}
The environmental interference derivative for an NE subsystem is given by\footnote{The notation $I_\kappa({\sf NE}_{-\kappa}({\sf p}_\kappa))[n]$ illustrates that the interference seen by user $\kappa$, $I_\kappa[n]$, is a result of the power allocation from other users, ${\sf NE}_{-\kappa}$, which is caused by the power allocation of user $\kappa$, ${\sf p}_\kappa$.}
\begin{equation}\label{eqn:varpi-ne2}
\frac{\partial I_\kappa({\sf NE}_{-\kappa}({\sf p}_\kappa))[n]}{\partial p_\kappa[n]}=-\sum_{i=1\atop i\ne\kappa}^K\theta_{\kappa i}[n]\theta_{i\kappa}[n]{\sf sgn}(p_i[n])\le 0,
\end{equation}
where ${\sf NE}_{-\kappa}({\sf p}_\kappa)\triangleq\{{\sf NE}_1({\sf p}_\kappa),\dots,{\sf NE}_{\kappa-1}({\sf p}_\kappa),{\sf NE}_{\kappa+1}({\sf p}_\kappa),\dots,{\sf NE}_K({\sf p}_\kappa)\}$.
\end{lemma}

\proof Noting that $I_\kappa[n]=\sum_{i\ne\kappa}p_i[n]\theta_{i\kappa}[n]=\sum_{i\ne\kappa}NE_i({\sf p}_\kappa)[n]\theta_{i\kappa}[n]$, we have
\begin{equation}
\frac{\partial I_\kappa({\sf NE}_{-\kappa}({\sf p}_\kappa))[n]}{\partial p_\kappa[n]}=\frac{\partial\left(\sum_{i\ne\kappa}NE_i({\sf p}_\kappa)[n]\theta_{i\kappa}[n]\right)}{\partial p_\kappa[n]}=\sum_{i=1\atop i\ne\kappa}^K\theta_{i\kappa}[n]\frac{\partial NE_i({\sf p}_\kappa)[n]}{\partial p_\kappa[n]}.
\end{equation}
Then, the final result can be obtained by substituting the result of Lemma 1.\endproof

\begin{theorem}
NE is unique in the OFDMA game if the environmental interference derivative satisfies
\begin{equation}\label{eid-ne}
\varphi_k[n]=\frac{\partial I_k({\sf NE}_{-k}({\sf p}_k))[n]}{\partial p_k[n]}>-1,~\mbox{for all ${\sf p}_k$ and for all $k,n$}.
\end{equation}
\end{theorem}

\proof To prove the result, it suffices to show that $w_k^*$ is unique for all $k$, and we are not interested in those $p_k^*[n]=0$ because they do not cause interference to others nor consume any power budget.

At NE, for $p_k^*[n]>0$, we have
\begin{equation}
w_k^*=p_k^*[n]+c_k^*[n].
\end{equation}
Then, the relationship between the water-level and the optimal power allocation can be studied by differentiating $w_k^*$ with respect to $p_k^*[n]$ which gives $\frac{\partial w_k^*}{\partial p_k^*[n]}=1+\varphi_k^*[n]$. If (\ref{eid-ne}) is true, or $\varphi_k^*[n]>-1$, then $\frac{\partial w_k^*}{\partial p_k^*[n]}>0$. In other words, if we have two water-levels such that $w_k<\tilde{w}_k$, then their respective power allocations will satisfy $\sum_np_k[n]<\sum_n\tilde{p}_k[n]$. Hence, there is a unique $w_k^*$ such that $\sum_np_k^*[n]=P_k$.\endproof

\begin{Corollary}
Denoting $\bar\theta=\max_{i,k}\theta_{ki}$, a sufficient condition for the uniqueness of NE is
\begin{equation}\label{eqn:bartheta2}
\bar{\theta}<\frac{1}{\sqrt{K-1}}.
\end{equation}
\end{Corollary}

\proof If (\ref{eqn:bartheta2}) is true, then $1>(K-1)\bar{\theta}^2>\sum_{i=1\atop i\ne\kappa}^K\theta_{\kappa i}[n]\theta_{i\kappa}[n]{\sf sgn}(p_i[n])$, which implies that $\varphi_k[n]=-\sum_{i=1\atop i\ne\kappa}^K\theta_{\kappa i}[n]\theta_{i\kappa}[n]{\sf sgn}(p_i[n])>-1$. As a result, according to Theorem 1, NE is unique.\endproof

Theorem 1 provides a new way for analyzing the uniqueness of the equilibrium of an OFDMA game through the environmental interference derivative. On the other hand, comparing to the literature requiring $\bar\theta<\frac{1}{K-1}$ \cite{Yu-02,Scutari-07,Chung-03,Huang-06}, Corollary 1 provides a less stringent sufficient condition for the uniqueness of NE.

\subsection*{B. OSE and A Free Game Implementation}
\begin{definition}
For OSE where user $\kappa$ is a foresighted user and the rest are myopic followers, we have
\begin{equation}\label{ose}
\left\{\begin{aligned}
R_\kappa({\sf p}_\kappa^*,{\sf NE}_{-\kappa}({\sf p}_\kappa^*))&\ge R_\kappa({\sf p}_\kappa,{\sf NE}_{-\kappa}({\sf p}_\kappa)),\\
R_k({\sf p}_k^*,{\sf p}_{-k}^*({\sf p}_k^*))&\ge R_k({\sf p}_k,{\sf p}_{-k}^*({\sf p}_k^*))~\forall k\ne\kappa.
\end{aligned}\right.
\end{equation}
\end{definition}

The OSE above, as defined in the conventional fashion, is by nature a bi-level game for the foresighted user $\kappa$ because the action user $\kappa$ takes is expecting an NE response over the rest of the users, as seen in (\ref{ose}). Clearly, the non-Stackelberg users $(k\ne\kappa)$ are myopic and play the game as in NE. In the following, we will study the environmental property of OSE and use it to create a free game version of OSE.

\begin{theorem}
Given the foresighted user $\kappa$ in OSE, we have
\begin{equation}\label{eqn:eid-ose}
\varphi_\kappa^*[n]\ge-\frac{c_\kappa^*[n]}{2c_\kappa^*[n]+p_\kappa^*[n]}.
\end{equation}
\end{theorem}

\proof Given that at OSE, we have ${\sf p}_\kappa^*$, we now consider a power allocation strategy, $\tilde{\sf p}_\kappa$, such that $\tilde{p}_\kappa[l]=p^*_\kappa[l]~\forall l\ne m,n$, but $\tilde{p}_\kappa[n]=p_\kappa^*[n]+\Delta p$ and $\tilde{p}_\kappa[m]=p_\kappa^*[m]-\Delta p$ for some $1\le m\ne n\le N$ and small $\Delta p>0$. Denote the interference patterns for ${\sf p}_\kappa^*$ and $\tilde{\sf p}_\kappa$, respectively, as ${\sf I}_\kappa$ and $\tilde{\sf I}_\kappa$. Using Lemma 2, we can write $\tilde{I}_\kappa[n]=I_\kappa[n]-\Delta I[n]$ and $\tilde{I}_\kappa[m]=I_\kappa[m]+\Delta I[m]$ for some small $\Delta I[n]$ and $\Delta I[m]>0$. As user $\kappa$'s rate is maximized by ${\sf p}^*_\kappa$ at OSE, we therefore have $R_\kappa(\tilde{\sf p}_\kappa)-R_\kappa({\sf p}_\kappa^*)\le 0$ implying
\begin{multline}
\log_2\left(1 + \frac{{p_\kappa^*[n]+\Delta p}}{{\sigma_\kappa[n] + I_\kappa^*[n]-\Delta I[n]}}\right)-\log_2\left(1 + \frac{{p_\kappa^*[n]}}{{\sigma_\kappa[n]+ I_\kappa^*[n]}}\right)\\
+\log_2\left(1 + \frac{{p_\kappa^*[m]-\Delta p}}{{\sigma_\kappa[m] + I_\kappa^*[m] +\Delta I[m]}}\right)-\log_2\left(1 + \frac{{p_\kappa^*[m]}}{{\sigma_\kappa[m]+ I_\kappa^*[m]}}\right)\le 0.
\end{multline}
This can further be simplified to
\begin{multline}\label{eqn:th3-1}
\left(\frac{\sigma_\kappa[n]+ I_\kappa^*[n]-\Delta I[n]+p_\kappa^*[n]+\Delta p}{\sigma_\kappa[n]+ I_\kappa^*[n]-\Delta I[n]}\right)\left(\frac{\sigma_\kappa[n]+ I_\kappa^*[n]}{\sigma_\kappa[n]+I_\kappa^*[n]+p_\kappa^*[n]}\right)\\
\times\left(\frac{\sigma_\kappa[m]+I_\kappa^*[m]+\Delta I[m]+p_\kappa^*[m]-\Delta p}{\sigma_\kappa[m]+ I_\kappa^*[m]+\Delta I[m]}\right)\left(\frac{\sigma_\kappa[m]+
I_\kappa^*[m]}{\sigma_\kappa[m]+I_\kappa^*[m]+p_\kappa^*[m]}\right)\le 1.
\end{multline}
Now, using (\ref{eqn:th3-1}) and noting that it is valid also after swapping $m$ and $n$ because subchannel indices $m$ and $n$ are arbitrary, it can be easily shown after some manipulations that
\begin{multline}
\left(1-\frac{(\Delta I[n]-\Delta p)^2}{(\sigma_\kappa[n]+I_\kappa^*[n]+p_\kappa^*[n])^2}\right)\left(1-\frac{(\Delta I[m]-\Delta p)^2}{(\sigma_\kappa[m]+I_\kappa^*[m]+p_\kappa^*[m])^2}\right)\\
\le\left(1-\frac{\Delta I[n]^2}{(\sigma_\kappa[n]+ I_\kappa^*[n])^2}\right)\left(1-\frac{\Delta I[m]^2}{(\sigma_\kappa[m]+I_\kappa^*[m])^2}\right)~\forall m,n,
\end{multline}
\begin{equation}\label{seip}
\Rightarrow\frac{{\left| {\Delta p-\Delta I[n]}\right|}}{{\underbrace {\sigma_\kappa[n]+I_\kappa^*[n]}_{c_\kappa^*[n]}+p_\kappa^*[n]}}\ge\frac{{\Delta I[n]}}{{\sigma_\kappa[n]+I_\kappa^*[n]}} \Rightarrow\left|\frac{\Delta I[n]}{\Delta p}\right|\le\frac{c_\kappa^*[n]}{2c_\kappa^*[n]+p_\kappa^*[n]}~\forall n.
\end{equation}
Taking the limit $\Delta p\to 0$ and knowing that $\varphi_\kappa^*[n]\le 0$ (Lemma 2), we get the desired result of (\ref{eqn:eid-ose}).\endproof

\begin{theorem}
As long as the environmental interference derivative exists, $\varphi_\kappa[n]$, OSE is unique.
\end{theorem}

\proof From our GWF analysis, for $p_\kappa^*[n]>0$, we have $w_\kappa^*=\eta_\kappa^*[n]+p_\kappa^*[n]$ which gives
\begin{equation}
w=\frac{c^2+\varphi p^2}{c-\varphi p}+p
\end{equation}
where the indices $\kappa$ and $n$ as well as the superscript $*$ are omitted for convenience. In \cite[Proposition 1]{Su-09TSP}, it is known that in OSE, $\varphi$ is a constant and hence $\frac{\partial\varphi}{\partial p}=0$. Then, $\frac{\partial w}{\partial p}$ can be derived as
\begin{align}
\frac{\partial w}{\partial p}&=\frac{(c-\varphi p)(2c\varphi+2p\varphi)-(c^2+\varphi p^2)(\varphi-\varphi)}{(c-\varphi p)^2}+1\\
&=\frac{2c\varphi+\varphi p+c}{c-\varphi p}.
\end{align}
Because $\varphi\le 0$ (Lemma 2), we have $c-\varphi p>0$. From Theorem 2, we also have $\varphi\ge-\frac{c}{2c+p}$. Therefore,
\begin{equation}
\frac{\partial w}{\partial p}\ge\frac{\left(-\frac{c}{2c+p}\right)(2c+p)+c}{>0}=0.
\end{equation}
Using a similar argument as in the proof of Theorem 1, $w_\kappa^*$ is unique and so does OSE.\endproof

Also, from our GWF formulation in Section II-C, we know that $\frac{\partial{\cal L}}{\partial p_\kappa^*[n]}=\frac{1}{w_\kappa^*}-\lambda$ which further gives
\begin{equation}\label{eqn:d2Ldp2}
\frac{\partial^2{\cal L}}{\partial (p_\kappa^*[n])^2}=-\left(\frac{1}{w_\kappa^*}\right)^2\frac{\partial w_\kappa^*}{\partial p_\kappa^*[n]}.
\end{equation}
From the proof of Theorem 3, we have already known that $\frac{\partial w_\kappa^*}{\partial p_\kappa^*[n]}\ge 0$. As a consequence, $\frac{\partial^2{\cal L}}{\partial (p_\kappa^*[n])^2}\le 0$, which is a sufficient condition for optimality of $p_\kappa^*[n]$ for maximizing user $\kappa$'s achievable rate.

The environmental interference derivative property of OSE in Theorem 2 is key to developing a free OSE game for self-optimization of OFDMA. To do so, we first create an artificial OFDMA game where user $\kappa$ is foresighted and other users are myopic, and the environmental interference derivative, $\phi$, meets
\begin{equation}
\phi_\kappa^*[n]=-\frac{1}{2+\frac{p_\kappa^*[n]}{c_\kappa^*[n]}}\le\varphi_\kappa^*[n]\le 0.
\end{equation}
The above artificial game will approach the bi-level OSE game if $\frac{p_\kappa^*[n]}{c_\kappa^*[n]}$ is large for all $n$ such that $p_\kappa^*[n]>0$ because $\phi_\kappa^*[n]\to\varphi_\kappa^*[n]$. In this case, the optimal power allocation for both games will be identical. We conjecture that for $\frac{N}{K}>1$, there will be a strong tendency that foresighted user $\kappa$ will {\em own} the subcarriers that it chooses to occupy and in so doing on those chosen subcarriers $\frac{p_\kappa^*[n]}{c_\kappa^*[n]}$ will be large and as a result the artificial game provides an accurate representation of the original bi-level OSE game.

Inspired by this, we propose to use the GWF by choosing the operational interference derivative (i.e., the belief of the environmental interference derivative) for foresighted user $\kappa$ as (see also Section II-C)
\begin{equation}\label{eqn:tildevarphi-ose}
\tilde{\varphi}_\kappa^*[n]=-\frac{c_\kappa^*[n]}{2c_\kappa^*[n]+p_\kappa^*[n]},~\mbox{or operationally}, \tilde{\varphi}_\kappa^t[n]=-\frac{c_\kappa^t[n]}{2c_\kappa^t[n]+p_\kappa^{t-1}[n]}.
\end{equation}
Therefore, we have the following free game for approaching the OSE.

\begin{proposition}
A free OSE game lets users iteratively and simultaneously optimize their strategies using GWF with user $\kappa$ choosing (\ref{eqn:tildevarphi-ose}) and other myopic users choosing $\tilde{\varphi}_k[n]=0~\forall k\ne\kappa$ so that
\begin{equation}\label{eqn:pkWF-ose}
\left\{\begin{aligned}
p_\kappa^{t}[n]&=\left(w_\kappa^t-\frac{\left(c_\kappa^t[n]\right)^2+\tilde{\varphi}^t_\kappa[n]\left(p_\kappa^{t-1}[n]\right)^2}{c_\kappa^t[n]
-\tilde{\varphi}^t_\kappa[n]p_\kappa^{t-1}[n]}\right)^+,~\mbox{where $\tilde{\varphi}_\kappa^t[n]$ in (\ref{eqn:tildevarphi-ose})},\\
p_k^{t}[n]&=\left(w_k^t-c_k^t[n]\right)^+,~\mbox{for }k\ne\kappa.
\end{aligned}\right.
\end{equation}
\end{proposition}

With the operational interference derivative (\ref{eqn:tildevarphi-ose}), user $\kappa$ in this free game does not need to wait for an NE response of the rest of the users while adapting its power allocation strategy, and the equilibrium of the free game can be achieved by simultaneous GWF in any arbitrary order.

\begin{theorem}
If $\varphi_\kappa^*[n]=\frac{\partial I_\kappa^*[n]}{\partial p_\kappa^*[n]}>-0.5$, the free OSE game has a unique equilibrium and the GWF in (\ref{eqn:pkWF-ose}) achieves the unique equilibrium which is also the global optimum for maximizing user $\kappa$'s rate.
\end{theorem}

\proof With the operational interference derivative (\ref{eqn:tildevarphi-ose}) in GWF, we have, for $p_\kappa^*[n]>0$, that
\begin{equation}\label{eqn:ose-wcp}
w_\kappa^*=\frac{(c_\kappa^*[n])^2+c_\kappa^*[n]p_\kappa^*[n]}{c_\kappa^*[n]-\tilde{\varphi}_\kappa^*[n]p_\kappa^*[n]}=c_\kappa^*[n]+\frac{p_\kappa^*[n]}{2}.
\end{equation}
Therefore, if $\varphi_\kappa^*[n]>-0.5$, then we get
\begin{equation}
\frac{\partial w_\kappa^*}{\partial p_\kappa^*[n]}=\varphi_\kappa^*[n]+\frac{1}{2}\ge 0.
\end{equation}
Since $\frac{\partial w_\kappa^*}{\partial p_\kappa^*[n]}\ge 0$ implies that $w_k^*$ is unique, the equilibrium of the free game is unique. For the same reason, as shown in (\ref{eqn:d2Ldp2}) previously, we have $\frac{\partial^2{\cal L}}{\partial (p_\kappa^*[n])^2}\le 0$ and therefore, $p_\kappa^*[n]$ obtained by the GWF in (\ref{eqn:pkWF-ose}) will converge to the unique equilibrium and is the global optimum under the condition $\varphi_\kappa^*[n]>-0.5$.\endproof

Unlike the bi-level game, in the free game, $\varphi_\kappa[n]=\frac{\partial I_\kappa[n]}{\partial p_\kappa[n]}\ne\frac{\partial I_\kappa({\sf NE}_{-\kappa}({\sf p}_\kappa))[n]}{\partial p_\kappa[n]}$ because the interference seen by user $\kappa$ is no longer the outcome of the NE response of all the myopic users. Also, within the free game, it does not seem possible to know the exact value for $\varphi_\kappa[n]=\frac{\partial I_\kappa[n]}{\partial p_\kappa[n]}$. However, the way the free game is carried out is based on the belief that the interference derivative has the property (\ref{eqn:pkWF-ose}) which is the property that permits the free game to approach the bi-level OSE. In this belief, $\tilde{\varphi}_\kappa[n]>-0.5$.

\subsection*{C. ASE, Its Variants and A Free Game Implementation}
\begin{definition}
Users can all be completely foresighted and in this ASE, we have
\begin{equation}\label{ultimate-ase}
R_k({\sf p}_k^*,{\sf ASE}_{-k}({\sf p}_k^*))\ge R_k({\sf p}_k,{\sf ASE}_{-k}({\sf p}_k))~\forall k,
\end{equation}
where ${\sf ASE}_{-k}(\cdot)$ corresponds to the ASE strategies from all other users except user $k$ and is defined similarly as ${\sf NE}_{-k}(\cdot)$ used previously. Under this definition, ASE is a multilevel game with $(K-1)$ levels.
\end{definition}

Note that (\ref{ultimate-ase}) is the ultimate definition for ASE with complete foresightedness for all users. However, under (\ref{ultimate-ase}), users strategies are strongly coupled together and it is not known if this is achievable, if not impossible. This partly explains why in \cite{Xu-09} foresighted leaders are not supposed to be foresighted with each other and as a consequence in the case of ASE, users become myopic, losing their foresightedness capability. Motivated by our OSE results, instead, one could define ASE in the following way.

\begin{definition}
Users are all foresighted with the belief that others are myopic. In this ASE, we have
\begin{equation}\label{ne-ase}
R_k({\sf p}_k^*,\widetilde{\sf NE}_{-k}({\sf p}_k^*))\ge R_k({\sf p}_k,\widetilde{\sf NE}_{-k}({\sf p}_k))~\forall k,
\end{equation}
where $\widetilde{\sf NE}_{-k}$ indicates that user $k$ believes that other users will come to an NE in response to its action.
\end{definition}

In (\ref{ne-ase}), although users are not foresighted to each other, they are foresighted to some belief of the overall network response. In particular, every user has the belief that all other users are myopic and give an NE response to its action. This version of ASE is a natural extension to OSE and can be interpreted as having $K$ interfered OSE games. The main issue of (\ref{ne-ase}) is that the construction of the game is limited to ``foresighted-myopic'' interaction and the OSE games will interfere with each other, whereas in ASE one would expect to facilitate ``foresighted-foresighted'' interaction, which is the major challenge. To nurture the required ``foresighted-foresighted'' exchange, we have the following theorem.

\begin{theorem}
The environmental interference derivatives for ASE, $\{\varphi_k^*[n]\}$, satisfy
\begin{equation}\label{eqn:varpi-ase}
\varphi_k^*[n]=\sum_{i=1\atop i\ne k}^K\frac{\theta_{ki}[n]\theta_{ik}[n]{\sf sgn}(p_i^*[n])}{\varphi_i^*[n]}.
\end{equation}
\end{theorem}

\proof First, by definition, we have
\begin{equation}
\varphi_k^*[n]=\frac{\partial I_k^*[n]}{\partial p_k^*[n]}=\sum_{i=1\atop i\ne k}^K{\sf sgn}(p_i^*[n])\theta_{ik}[n]\frac{\partial p_i^*[n]}{\partial p_k^*[n]}=\sum_{i=1\atop i\ne k}^K{\sf sgn}(p_i^*[n])\theta_{ik}[n]\frac{\partial p_i^*[n]}{\partial I_i^*[n]}\frac{\partial I_i^*[n]}{\partial p_k^*[n]},
\end{equation}
in which $\frac{\partial p_i^*[n]}{\partial I_i^*[n]}=\left(\frac{\partial I_i^*[n]}{\partial p_i^*[n]}\right)^{-1}=(\varphi_i^*[n])^{-1}$, and $\frac{\partial I_i^*[n]}{\partial p_k^*[n]}=\theta_{ki}[n]$. As such, the property (\ref{eqn:varpi-ase}) is obtained.\endproof


Theorem 5 provides a key property of ASE which facilitates the exchanges of foresightedness between users. With this result, Definition 3 in (\ref{ne-ase}) can go beyond a group of interfered OSEs to become a group of foresighted OSEs. We will show that it is possible to gain insight from the property (\ref{eqn:varpi-ase}) so that self-optimization OFDMA can be achieved. Before we address this, the following lemmas are useful.

%

\begin{lemma}
With user $k$ believing that other users $(\ell\ne k)$ are myopic, we have
\begin{equation}\label{eqn:ase-property}
\sum_{\ell\ne k}\theta_{\ell k}[n]\theta_{k\ell}[n]{\sf sgn}(p_\ell^*[n])\le\frac{c_k^*[n]}{2c_k^*[n]+p_k^*[n]}.
\end{equation}
\end{lemma}

\proof Based on the belief of user $k$, we start by writing $\frac{\Delta I_k^*[n]}{\Delta p_k^*[n]}$ as
\begin{equation}\label{eqn:lemma4-step}
\frac{\Delta I_k^*[n]}{\Delta p_k^*[n]}\stackrel{(a)}{=}\sum_{\ell\ne k}\theta_{\ell k}[n]\frac{\Delta p_\ell^*[n]}{\Delta I_\ell^*[n]}\frac{\Delta I_\ell^*[n]}{\Delta p_k^*[n]}{\sf sgn}(p_\ell^*[n])\stackrel{(b)}{\ge}-\frac{c_k^*[n]}{2c_k^*[n]+p_k^*[n]},
\end{equation}
where $(a)$ is by definition and $(b)$ is due to (\ref{eqn:eid-ose}) (true as long as the rate at the equilibrium is the highest). Now, knowing $\frac{\Delta p_\ell^*[n]}{\Delta I_\ell^*[n]}=-1$ (as in the proof of Lemma 1) and $\frac{\Delta I_\ell^*[n]}{\Delta p_k^*[n]}=\theta_{k\ell}$ yield the result (\ref{eqn:ase-property}).\endproof

\begin{proposition}
For a self-optimized OFDMA network where each user knows only its local CSI, the environmental interference derivative for user $k$ to achieve ASE (Definition 3) should satisfy
\begin{equation}\label{eqn:varpi-ase-ofdma}
\varphi_k^*[n]\ge-\sqrt{\frac{c_k^*[n]}{2c_k^*[n]+p_k^*[n]}}~\forall k,n.
\end{equation}
\end{proposition}

\proof Note that (\ref{eqn:varpi-ase}) is true only if $p_k^*[n]>0$ because otherwise the subchannel is not used and $p_k^*[n]=0$. Therefore, more precisely, if $\theta_{ij}[n]\triangleq 0$ for $i= j$, then we have
\begin{subequations}
\begin{align}
\sum_{k=1}^K\varphi_k^*[n]&\stackrel{(a)}{=}\sum_{k=1}^K\frac{{\sf sgn}(p_k^*[n])}{\varphi_k^*[n]}\sum_{\ell=1}^K\theta_{k\ell}[n]\theta_{\ell k}[n]{\sf sgn}(p_\ell^*[n])\label{eqn:44a}\\
&\stackrel{(b)}{\le} \sum_{k=1}^K\frac{{\sf sgn}(p_k^*[n])}{\varphi_k^*[n]}\left(\frac{c_k^*[n]}{2c_k^*[n]+p_k^*[n]}\right),\label{eqn:ase-th2}
\end{align}
\end{subequations}
where $(a)$ is due to summing both sides of (\ref{eqn:varpi-ase}) over all $k$ and $(b)$ is due to the result in Lemma 3. With self-optimization in mind that each user should operate based on only its local CSI observation, i.e., $\{c_k[n]\}$ for user $k$, the above property can be enforced by comparing term-by-term on both sides giving rise to
\begin{equation}
\varphi_k^*[n]\le\frac{{\sf sgn}(p_k^*[n])}{\varphi_k^*[n]}\left(\frac{c_k^*[n]}{2c_k^*[n]+p_k^*[n]}\right)
\Rightarrow -\sqrt{\frac{c_k^*[n]}{2c_k^*[n]+p_k^*[n]}}\le\varphi_k^*[n]\le 0.
\end{equation}
Note that the negative is chosen to give (\ref{eqn:varpi-ase-ofdma}) because $\varphi_k^*[n]\le 0$.\endproof

It is worth emphasizing that (\ref{eqn:ase-th2}) and hence (\ref{eqn:varpi-ase-ofdma}) is a {\em weaker} environmental interference derivative property for ASE which is a result deduced from the original property (\ref{eqn:varpi-ase}) but satisfying (\ref{eqn:varpi-ase-ofdma}) does not necessarily imply (\ref{eqn:varpi-ase}). Nonetheless, it does bring the possibility of network foresightedness based on local channel observation and will therefore make self-optimization realizable by the following proposition.

\begin{proposition}
A free ASE game can be constructed by simultaneous GWF such that
\begin{equation}\label{eqn:pkWF-ase}
p_k^{t}[n]=\left(w_k^t-\frac{\left(c_k^t[n]\right)^2+\tilde{\varphi}^t_k[n]\left(p_k^{t-1}[n]\right)^2}{c_k^t[n]-\tilde{\varphi}^t_k[n]p_k^{t-1}[n]}\right)^+,~\mbox{where }\tilde{\varphi}_k^t[n]=-\sqrt{\frac{c_k^t[n]}{2c_k^t[n]+p_k^{t-1}[n]}}~\forall k.
\end{equation}
\end{proposition}

The above free ASE game proposed follows the same rationale behind the free OSE game described before. In particular, the operational interference derivative for carrying out GWF is chosen to satisfy the environmental interference derivative property of ASE (\ref{eqn:varpi-ase-ofdma}) with equality on the basis of approaching the original multilevel ASE at high SNRs. Furthermore, being a free game, the equilibrium can be achieved by simultaneous GWF in any order. The following theorem addresses the uniqueness of the free game.

\begin{theorem}
If $\varphi_k^*[n]>-\frac{c_k^*[n]}{4c_k^*[n]+p_k^*[n]}~\forall k,n$ such that $p_k^*[n]>0$, the free ASE game has a unique equilibrium and the GWF in (\ref{eqn:pkWF-ase}) achieves the unique equilibrium and is also optimal for rate maximization.
\end{theorem}

\proof To begin, we find it useful to define $\gamma_k^*[n]=\sqrt{\frac{2c_k^*[n]+p_k^*[n]}{c_k^*[n]}}$ so that $\tilde{\varphi}_k^*[n]=-\frac{1}{\gamma_k^*[n]}$ due to the GWF in (\ref{eqn:pkWF-ase}) and $p_k^*[n]=(\gamma_k^*[n])^2c_k^*[n]-2c_k^*[n]$. Then, for those $k,n$ such that $p_k^*[n]>0$, we have
\begin{align}
w_k^*[n]&=\frac{(c_k^*[n])^2+c_k^*[n]p_k^*[n]}{c_k^*[n]-\tilde{\varphi}_k^*[n]p_k^*[n]}
=\frac{c_k^*[n]\gamma_k^*[n](\gamma_k^*[n]+1)}{\gamma_k^*[n]+2}.
\end{align}
To see how $w_k^*[n]$ varies with respect to $p_k^*[n]$, we first obtain
\begin{equation}
\frac{\partial\gamma_k^*[n]}{\partial p_k^*[n]}=\frac{1}{2}\frac{1}{\sqrt{2+\frac{p_k^*[n]}{c_k^*[n]}}}\frac{\partial\left(\frac{p_k^*[n]}{c_k^*[n]}\right)}{\partial p_k^*[n]}=
\frac{1}{2}\frac{1}{\sqrt{2+\frac{p_k^*[n]}{c_k^*[n]}}}\frac{1}{c_k^*[n]}\left(1-\frac{p_k^*[n]\varphi_k^*[n]}{c_k^*[n]}\right)\ge 0 ~~(\because\varphi_k^*[n]\le 0).
\end{equation}
Therefore, $\gamma_k^*[n]$ increases as $p_k^*[n]$ increases. On the other hand, 
$\frac{\partial\left(\frac{\gamma_k^*[n]+1}{\gamma_k^*[n]+2}\right)}{\partial\gamma_k^*[n]}=\frac{1}{(\gamma_k^*[n]+2)^2}>0$
and hence if $p_k^*[n]$ increases, then $\gamma_k^*[n]$ will increase and therefore $\frac{\gamma_k^*[n]+1}{\gamma_k^*[n]+2}$ will increase accordingly.

It remains to show that $c_k^*[n]\gamma_k^*[n]$ is also increasing with $p_k^*[n]$. To do so, we note that $c_k^*[n]\gamma_k^*[n]=\sqrt{2(c_k^*[n])^2+p_k^*[n]c_k^*[n]}$. As a consequence, if $\varphi_k^*[n]>-\frac{c_k^*[n]}{4c_k^*[n]+p_k^*[n]}$ is satisfied, we obtain
\begin{equation}
\frac{\partial c_k^*[n]\gamma_k^*[n]}{\partial p_k^*[n]}=\frac{1}{2}\frac{1}{\sqrt{2(c_k^*[n])^2+p_k^*[n]c_k^*[n]}}\left[(4c_k^*[n]+p_k^*[n])\varphi_k^*[n]+c_k^*[n]\right]>0.
\end{equation}

As a result, under the aforementioned condition, $w_k^*[n]$ is increasing with $p_k^*[n]$ and hence the power constraint $\sum_np_k^*[n]$. Following the same argument as before, $w_k^*$ and hence the free ASE game is unique. From (\ref{eqn:d2Ldp2}), the equilibrium is also optimal in maximizing the user's rate, which completes the proof.\endproof

\section*{\sc IV. Convergence Analysis}
The power-interference game can be generally analyzed by recognizing that
\begin{equation}\label{ca0}
\left\{\begin{aligned}
\Delta p_k^{t}&=\rho_k^t\Delta I_k^t,~\mbox{for some }\rho_k^t\le 0~~\mbox{(observe and action; GWF reoptimization)},\\
\Delta I_k^{t+1}&=\varphi_k^t\Delta p_k^t,~\mbox{for some }\varphi_k^t\le 0~~\mbox{(action and reaction; environmental response)},
\end{aligned}\right.
\end{equation}
where the subcarrier index $n$ is omitted for convenience. In what follows, we can write
\begin{align}
\Delta I_k^{t+1}&=\varphi_k^t\Delta p_k^t=\sum_{i\ne k}{\Delta p_i^t\theta_{ik}},\\
\Delta p_k^{t+1}&=\rho_k^{t+1}\Delta I_k^{t+1}=\rho_k^{t+1}\varphi_k^t\Delta p_k^t=\rho_k^{t+1}\sum_{i\ne k}{\Delta p_i^t\theta_{ik}}.\label{eqn:dpdp}
\end{align}

Convergence can take place in two possible cases: (i) $|\Delta p_k^{t+1}|<|\Delta p_k^{t}|$ for all $t\ge t_0$ for some $t_0>0$, and (ii) $|\Delta p_k^{t+1}|>|\Delta p_k^{t}|$ for all $t\ge t_0$ for some $t_0>0$, the latter of which belongs to the case of strong interference channel. To proceed, we define $|\overline{\Delta p}^t|\triangleq\max_{k}|\Delta p_k^t|$, $|\underline{\Delta p}^t|\triangleq\min_{k}|\Delta p_k^t|$, $|\overline{\rho}^t|\triangleq\max_{k}|\rho_k^t|$, $|\underline{\rho}^t|\triangleq\min_{k}|\rho_k^t|$, $|\overline{\theta}|\triangleq\max_{i\ne j}|\theta_{ij}|$ and $|\underline{\theta}|\triangleq\min_{i\ne j}|\theta_{ij}|$. Our objective here is to illustrate that it is possible to obtain sufficient conditions for convergence on the channel parameters $|\overline{\theta}|$ and $|\underline{\theta}|$.

To have (i), we start by expressing
\begin{equation}
|\overline{\Delta p^{t+1}}|\stackrel{(a)}{=}\max_k\left|\rho_k^{t+1}\sum_{i\ne k}{\Delta p_i^{t}\theta_{ik}}\right|\le|\overline{\rho^{t+1}}||\overline{\Delta p^{t}}|\overline{\theta}(K-1),
\end{equation}
where $(a)$ is due to (\ref{eqn:dpdp}). Therefore, a sufficient condition for convergence is to have
\begin{equation}
|\overline{\rho^{t+1}}|\overline{\theta}(K-1)<1\Rightarrow \overline{\theta}<\frac{1}{(K-1)|\overline{\rho^{t+1}}|}
\end{equation}
because this will guarantee $|\overline{\Delta p^{t+1}}|<|\overline{\Delta p^{t}}|~\forall k$, leading to $\Delta p^\infty_k=0~\forall k$.

\begin{figure}[]
\begin{center}
\includegraphics[width=12cm]{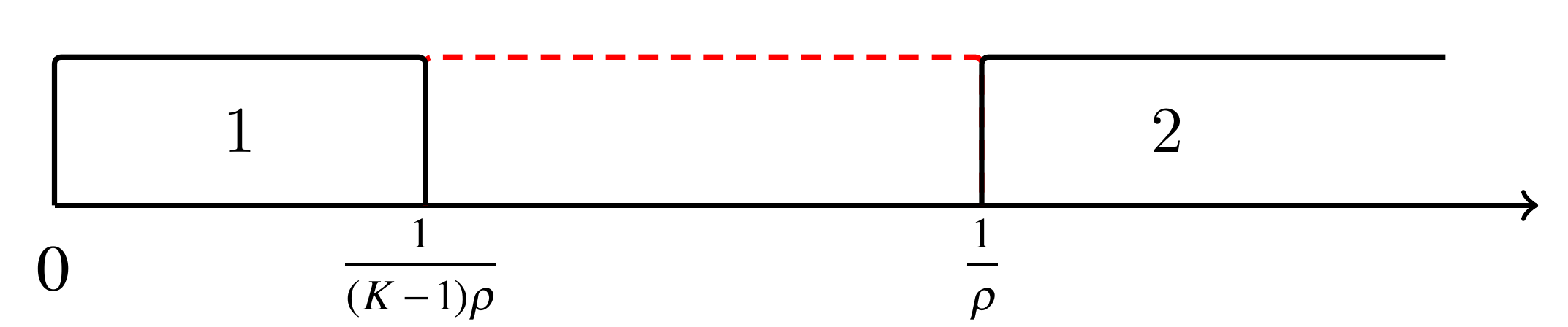}\caption{The sufficient convergence regions.}\label{fig_cr}
\end{center}
\end{figure}

On the other hand, for (ii) $|\Delta p_k^{t+1}|>|\Delta p_k^{t}|$, this will converge as well since $|\Delta p_k^t|$ is finite due to the total power constraint, $P_k$. This can be interpreted as the scenario where a user, say $k$, decides to allocate all its power to given subcarriers (if $\Delta p_k^t>0$) or to withdraw all its power (if $\Delta p_k^t<0$). The former illustrates the case that user $k$ wants to own the subcarriers and other users will have a strong tendency to leave the subcarriers due to severe interference caused by user $k$. A natural consequence is that those subcarriers will be occupied by user $k$ {\em only}, and hence, $\varphi^t_k=0$, resulting the strong interference channel studied in \cite{Maric-05,Kramer-04}. To obtain a sufficient condition for such convergence, we consider
\begin{equation}
|\Delta p_k^{t+1}|=\left|\rho_k^{t+1}\sum_{i\ne k}{\Delta p_i^{t}\theta_{ik}}\right|\ge |\underline{\rho^{t+1}}||\underline{\Delta p}^{t}|\underline\theta~\forall k,
\end{equation}
where we have used the fact that $\{\Delta p_i^{t}\}_{\forall i}$ are all of the same sign (all positive or all negative). As such, it will converge to a strong interference channel if
\begin{equation}
|\underline{\rho^{t+1}}|\underline\theta>1\Rightarrow \underline\theta>\frac{1}{|\underline{\rho^{t+1}}|}.
\end{equation}
Fig.~\ref{fig_cr} illustrates the convergence regions based on the two sufficient conditions above.

Based on the convergence regions described above, the convergence behaviors of various equilibriums can be analyzed by understanding $\rho_k^t$. In particular, we know that for $p_k^{t}>0$, we have
\begin{equation}\label{eqn:rhok}
\rho_k^t=\frac{\Delta p_k^{t}}{\Delta I_k^t}=\frac{\Delta (w_k^t-\eta_k^t)^+}{\Delta I_k^t}=-\frac{\Delta\eta_k^t}{\Delta I_k^t}=-\frac{\Delta\eta_k^t}{\Delta c_k^t}.
\end{equation}
As a result, we have the following lemma for evaluating $\rho_k^t$ of NE, OSE and ASE.

\begin{lemma}
Based on (\ref{eqn:rhok}), it can be derived that
\begin{equation}\label{ca1}
|\rho_k^t|=\left|\frac{\Delta\eta_k^t}{\Delta c_k^t}\right|=\left\{\begin{array}{cl}
1 & \mbox{if $\tilde\varphi_k^t=0$ (NE)},\\
1 & \mbox{if $\tilde\varphi_k^t=-\frac{c_k^t}{2c_k^t+p_k^{t-1}}$ (OSE with user $k$ as leader)},\\
\frac{1}{2}\sqrt{\frac{p_k^{t-1}}{c_k^t}} & \mbox{if $\tilde\varphi_k^t=-\sqrt{\frac{c_k^t}{2c_k^t+p_k^{t-1}}}$ when $\frac{p_k^{t-1}}{c_k^t}$ is large (ASE)}.
\end{array}\right.
\end{equation}
\end{lemma}

\begin{proof}
In this proof, for convenience, we will omit the subcarrier index $n$, the user index $k$ and the iteration index $t$. Also, we note that $|\rho|=\frac{\partial\eta}{\partial I}=\frac{\partial\eta}{\partial c}$. For NE, $\eta=c$ and therefore $|\rho|=1$.

In the case of OSE, using (\ref{eqn:etak}) and the corresponding definition of $\tilde\varphi$ in OSE, we have
\begin{equation}
\eta=\frac{c^2+\left(-\frac{c}{2c+p}\right)p^2}{c+\frac{c}{2c+p}p}=c-\frac{p}{2}~~\Rightarrow~~|\rho|=1.
\end{equation}

We start our analysis for the case of ASE by expressing
\begin{equation}\label{eqn:rho-ase}
\frac{\partial\eta}{\partial c}=\frac{c^2-2c\tilde\varphi p+c\frac{\partial\tilde\varphi}{\partial c}p^2-\tilde\varphi p^2+c^2\frac{\partial\tilde\varphi}{\partial c}p}{(c-\tilde\varphi p)^2}.
\end{equation}
On the other hand, we can also examine $\frac{\partial\tilde\varphi}{\partial c}$ to obtain
\begin{equation}
\frac{\partial\tilde\varphi}{\partial c}=\frac{1}{2\tilde\varphi}\frac{c^2}{(2c+p)^2}\frac{p}{c^2}=\frac{1}{2}\tilde\varphi^3\frac{p}{c^2}.
\end{equation}
Substituting this result into (\ref{eqn:rho-ase}) then gives
\begin{equation}
\frac{\partial\eta}{\partial c}=\frac{c^2-2c\tilde\varphi p+\frac{1}{2}\tilde\varphi^3\frac{p^3}{c}-\tilde\varphi p^2+\frac{1}{2}\tilde\varphi^3 p^2}{(c-\tilde\varphi p)^2}.
\end{equation}
Now, for $p\gg c$ or large $\frac{p}{c}$, we have $\tilde\varphi=-\sqrt{\frac{c}{2c+p}}\approx-\sqrt{\frac{c}{p}}$. Consequently, 
$|\rho|=\frac{\partial\eta}{\partial c}\approx\frac{1}{2}\tilde\varphi\frac{p}{c}=\frac{1}{2}\sqrt{\frac{p}{c}}$.
This completes the proof and knowing $|\rho|$ will be useful in understanding the convergence behavior.
\end{proof}

For NE, $|\rho^t_k|=1$ and a sufficient condition for convergence is $\overline\theta<\frac{1}{K-1}$, as reported in \cite{Yu-02,Scutari-07,Chung-03,Huang-06}. Also, our analysis shows that $\underline\theta>1$ will ensure convergence as well and in this case a given subcarrier will be occupied by one user only, which aligns with the result for the Gaussian strong interference channel in \cite{Maric-05,Kramer-04}. Since $|\rho^t_k|=1$ for OSE, it has the same convergence behavior as NE.

For ASE, as opposed to NE and OSE, $|\rho_k^t|$ can take very different values depending on $\frac{p_k^{t-1}}{c_k^t}$. Particularly, if $\frac{p_k^{t-1}}{c_k^t}$ is large, then $|\rho_k^t|$ will be very large and therefore convergence will tend to occur on the second region $\underline\theta\gtrsim 0$, in which case according to our analysis users at the equilibrium will be made orthogonal.

\section*{\sc V. Simulation Results}
In this section, we provide simulation results of the sum-rates achieved by the proposed algorithms. The iterative spectrum balancing (ISB) method in \cite{Yu-06}, which is a near-optimal {\em centralized} algorithm, is used as a performance upper bound. First, we consider an example 2-user 2-subcarrier network (i.e., $K=2,N=2$) \cite{Su-09TWC} where $\sigma_1[1]=\sigma_2[2]=4, \sigma_1[2]=\sigma_2[1]=1, \theta_{12}[1]=\theta_{12}[2]=\theta_{21}[1]=\theta_{21}[2]=0.5, P_1=P_2=10$ so that
\begin{eqnarray}
\left\{\begin{aligned}
R_1&=R_1[1]+R_1[2]=\log_2\left(1+\frac{p_1[1]}{4+0.5p_2[1]}\right)+\log_2\left(1+\frac{p_1[2]}{1+0.5p_2[2]}\right),\\
R_2&=R_2[1]+R_2[2]=\log_2\left(1+\frac{p_2[1]}{1+0.5p_1[1]}\right)+\log_2\left(1+\frac{p_2[2]}{4+0.5p_1[2]}\right).
\end{aligned}\right.
\end{eqnarray}
The solutions for NE, OSE and ASE (due to Definition 2 in (\ref{ultimate-ase})) can be easily derived analytically as
\begin{eqnarray}
\mbox{NE:}~~\left\{\begin{aligned}
p_1[1]&=2,\\
p_1[2]&=8,\\
p_2[1]&=8,\\
p_2[2]&=2,
\end{aligned}\right.~~
\mbox{OSE:}~~\left\{\begin{aligned}
p_1[1]&=0,\\
p_1[2]&=10,\\
p_2[1]&=9,\\
p_2[2]&=1,
\end{aligned}\right.~~\mbox{and}~~
\mbox{ASE:}~~\left\{\begin{aligned}
p_1[1]&=0,\\
p_1[2]&=10,\\
p_2[1]&=10,\\
p_2[2]&=0.
\end{aligned}\right.
\end{eqnarray}

The simulation results for this channel are given in Table 1, where it is observed that the proposed iterative GWF algorithm can be used to take users to the equilibria of interest. In particular, ASE tends to put the users into orthogonal subchannels and achieves the highest user rates.

\begin{figure}[]
\begin{center}
\includegraphics[width=8cm]{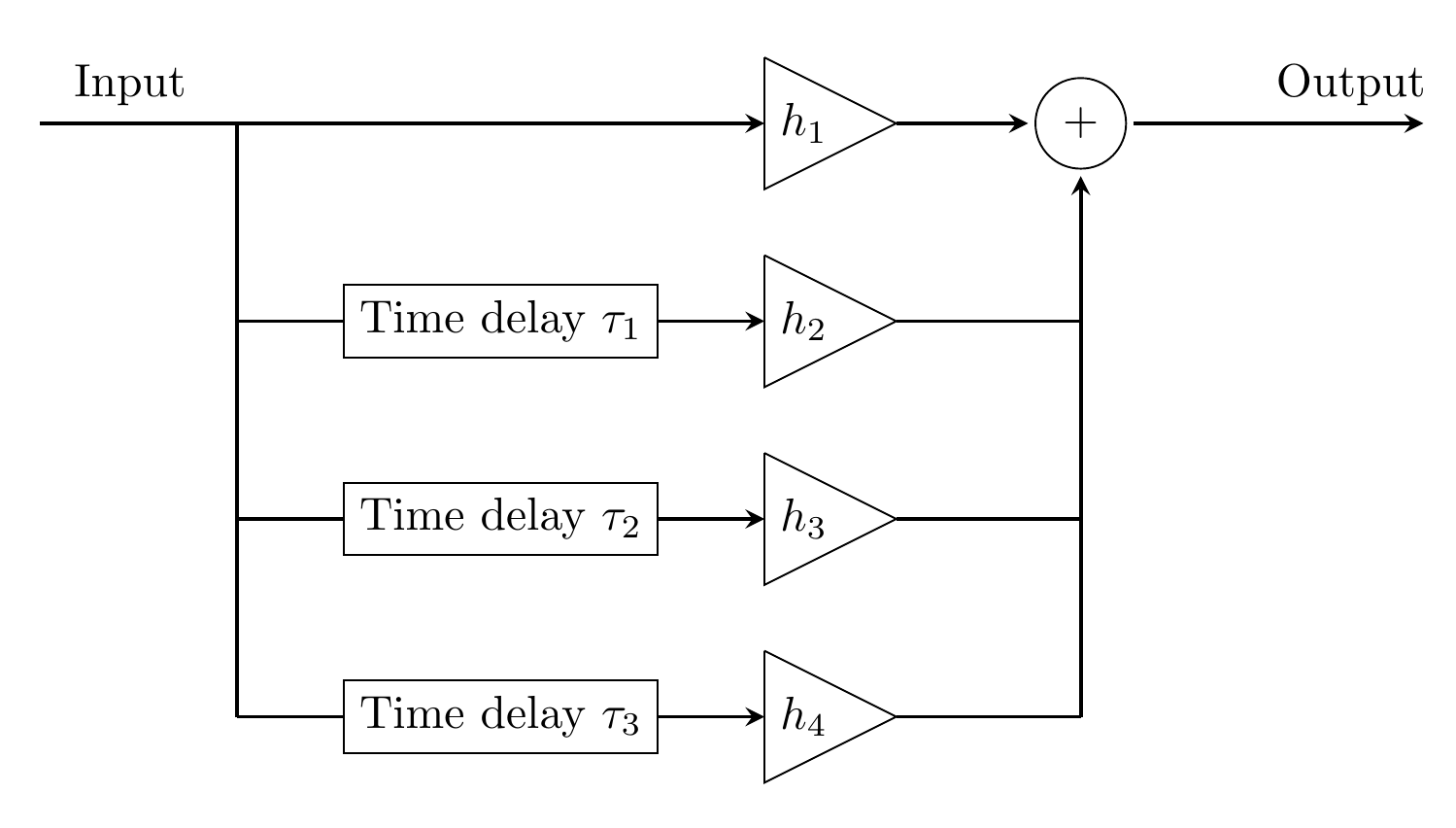}\caption{A four-ray Rayleigh fading channel model.}\label{fig_4rr}
\end{center}
\end{figure}

Next, we provide simulation results averaged over a large number of independent channel realizations. In the simulations, we model each subcarrier channel by an equal-power four-ray Rayleigh fading channel (see Fig.~\ref{fig_4rr}) \cite{Rappaport-96}, i.e.,
$|H_{ij}[n]|^2=|h^{(1)}_{ij}[n]|^2+|h^{(2)}_{ij}[n]|^2+|h^{(3)}_{ij}[n]|^2+|h^{(4)}_{ij}[n]|^2,~\forall i,j$,
and we assume that ${\tt E}[\left|H_{ij}[n]\right|^2]=x$ and ${\tt E}[|h^{(\ell)}_{ij}[n]|^2]=0.25x~\forall\ell$, for $i\ne j$. For $i=j$, we set ${\tt E}[\left|H_{kk}[n]\right|^2]=1$ and ${\tt E}[|h^{(\ell)}_{kk}[n]|^2]=0.25~\forall\ell$. Hence, $x$ is the parameter that measures the relative severeness of the interference channel.

Table 2 provides results showing the likelihood of convergence for different games with $P_k=100~\forall k$ and $N_k[n]=0.01~\forall k,n$. As shown, there is a small percentage of cases where NE and OSE diverge and as the number of users increases, this will become more problematic. In contrast, ASE always converges except when $x$ is small meaning that the channel has only very weak crosstalk. This is because for small $x$, noise becomes the dominant effect which does not react to the players' actions but users in ASE tend to mistakenly regard the overall noise as interference due to other users. As such, there will be a small percentage of times (less than $5\%$) for ASE to diverge. In terms of convergence speed, nevertheless, ASE generally would take longer to converge than NE and OSE (results not shown here due to space).

\begin{figure}[]
\begin{center}
\includegraphics[width=16cm]{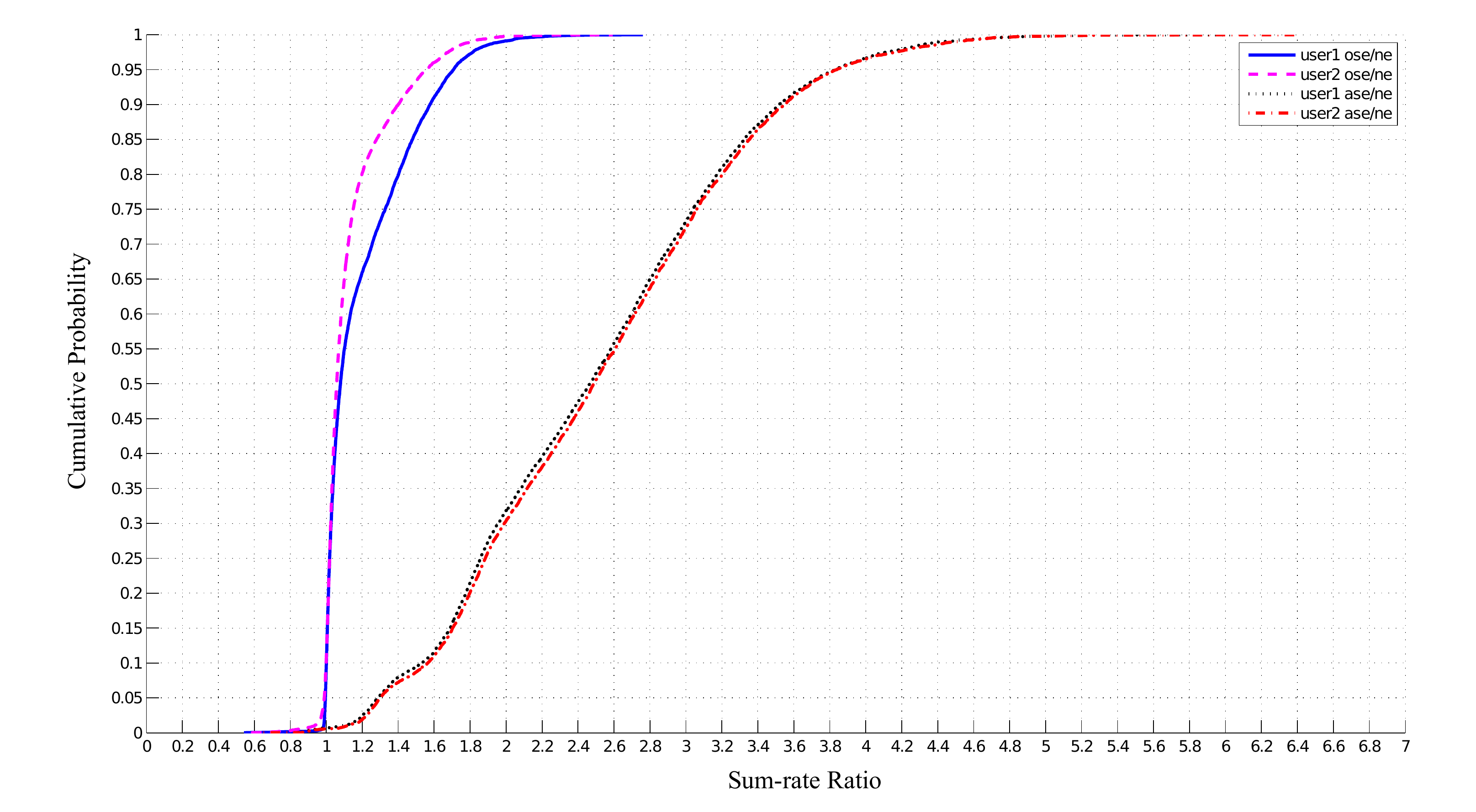}\caption{The cumulative density functions for the rate-ratios.}\label{fig_cdfs}
\end{center}
\end{figure}

In Table 3, results are provided for the average users' sum-rates for a $3$-user $9$-subcarrier interference channel, with $P_k=100~\forall k$, $N_k[n]=0.01~\forall k,n$ and $x=0.4$. Results show that there is a considerable gain in the sum-rates using ASE over NE and OSE. The cumulative density functions for the sum-rate ratios between ASE and NE users and between OSE and NE users are given in Fig.~\ref{fig_cdfs}. Results illustrate that for ASE it is possible to have the sum-rate $5$ times greater than what is achieved by NE. Furthermore, we compare the performance between ASE and ISB \cite{Yu-06}. Such results are given in Table 4 where we assumed $P_k=100~\forall k$, $N_k[n]=0.01~\forall k,n$ and $x=0.4$. Results show that the average per-user sum-rate for ASE is about $90\%$ of that for ISB and is not decreased much when the number of users/subcarriers increases.

\begin{figure}[]
\begin{center}
\includegraphics[width=14cm]{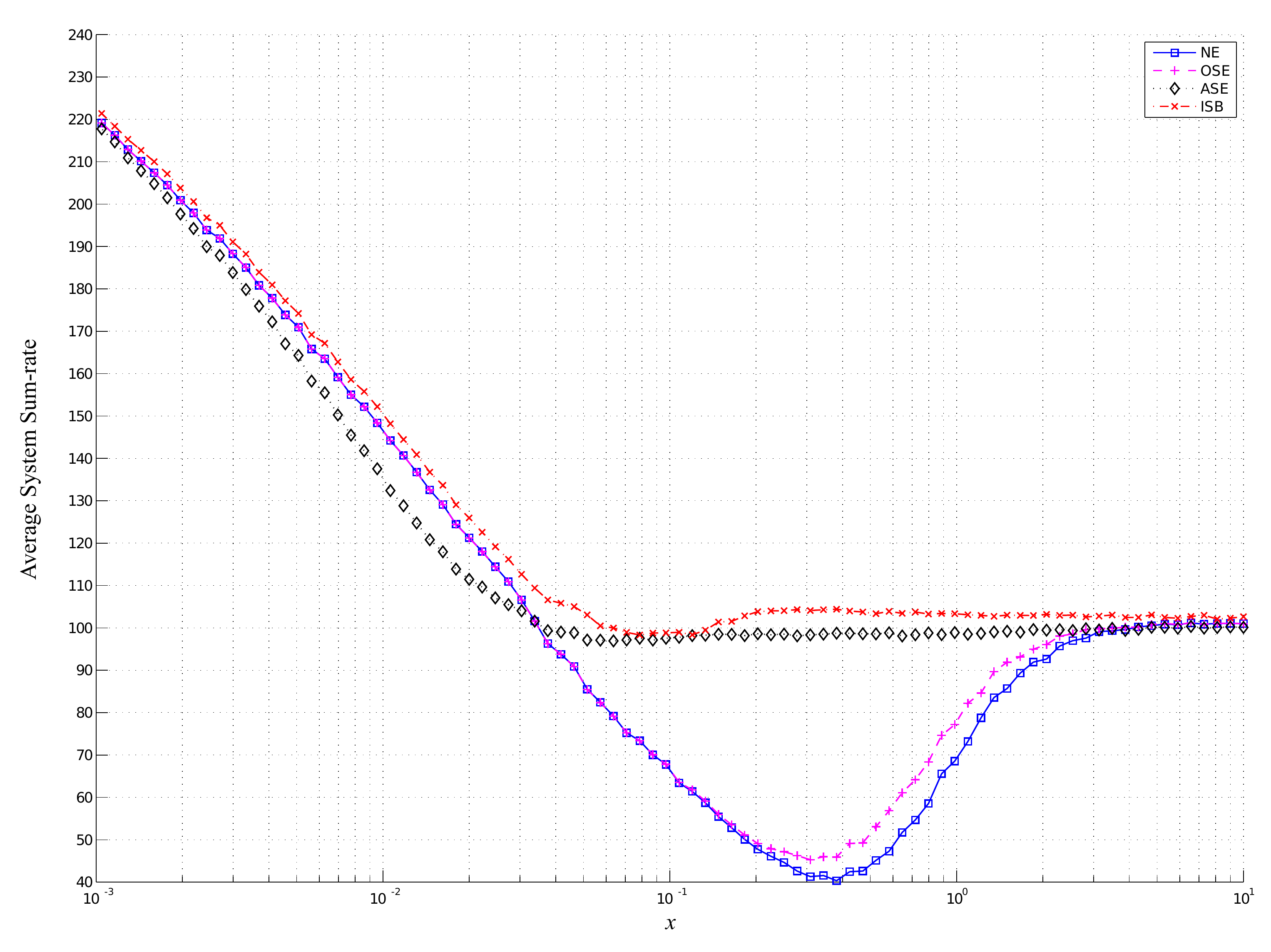}\caption{The average system sum-rate comparison for various $x$.}\label{fig_srate}
\end{center}
\end{figure}

To provide a more complete comparison between NE, OSE, ASE and ISB, in Fig.~\ref{fig_srate}, we provide the average system sum-rate results for various $x$ (which varies the severity of interference between users). A $3$-user $9$-subcarrier channel with $P_k=100~\forall k$, $N_k[n]=0.01~\forall k,n$ and $x$ from $0.001$ to $10$ is considered. We discuss the results by looking into three separate regions: (1) $x\le 0.03$, (2) $0.03<x\le 4$ and (3) $x>4$. In Region (1), $x$ is so small that the channel is reduced to $K$ parallel single-user OFDMA channels. In this case, users do not interfere with each other and NE, OSE and ASE all achieve sum-rate performance close to the centralized ISB, with ASE being slightly inferior than NE and OSE. Basically, all the game-theoretic approaches operate under an assumption that the players' strategies affect the interference patterns the users see but for small $x$, this is no longer true. For this reason, as ASE users are the most aggressive ones in playing, they are slightly inferior than NE and OSE users. From our convergence analysis in Section IV, it is understood that the superior performance of ASE is due to the possibility of having a large $\rho_k$ leading to convergence in region 2 in Fig.~\ref{fig_cr}. Using (\ref{ca1}), for ASE, we can roughly estimate
\begin{equation}
|\rho_k[n]|\approx \frac{1}{2}\sqrt{\frac{\frac{P_k}{N/K}}{N_k[n]}}=28.57.
\end{equation}
Therefore, for ASE to work well, we need to have $\underline\theta>\frac{1}{|\rho|}\approx 0.035$ (the dominant sufficient condition for ASE to converge), which agrees very much with the threshold of $x$ for Region (1).

Region (2) covers the most typical scenarios of interference channels. In this region, the performance differences between ISB and game-theoretic equilibria become much more significant. In particular, the sum-rates of NE and OSE are significantly compromised, while ASE is able to achieve the sum-rate close to that of ISB. Finally, in Region (3), NE, OSE and ASE all appear to converge to the same performance which is also very close to that for ISB. This is because when $x$ is very large (i.e., the interference links are much stronger than the desired links), this forces the gaming strategies to completely avoid users sharing any subcarriers \cite{Maric-05,Kramer-04}. Consequently, they all perform equally and as well as ISB.

\section*{\sc VI. Conclusion}
This paper studied the competition properties of an OFDMA channel. By introducing the environmental and operational interference derivatives, we devised iterative GWF algorithms that can take users to the ASE, by exploiting local CSI at the users, resulting self-optimized OFDMA. Results revealed that ASE achieves the average system sum-rate very close to that of ISB, a centralized near-optimal solution.

{\renewcommand{\baselinestretch}{1.1}
\begin{footnotesize}

\end{footnotesize}}

\begin{table}[]
\caption{Simulation results for the 2-user 2-subcarrier deterministic channel.}
\begin{center}
\begin{tabular}{c||c|c|c|c|c|c}
\hline Equilibrium&\multicolumn{2}{c|}{NE} &\multicolumn{2}{c|}{OSE}&\multicolumn{2}{c}{ASE}\\\hline\hline
User $k$ & User $1$ & User $2$ & User $1$ & User $2$ & User $1$ & User $2$\\\hline
$p_k[1]$ & $2.00$ & $8.00$ & $0$ & $9.00$ & $0$ & $9.99$\\\hline
$p_k[2]$ & $8.00$ & $2.00$ & $9.99$ & $1.00$ & $9.99$ & $0$\\\hline
Rate $R_k$ & $2.6439$ & $2.6438$ & $2.9386$ & $3.4739$ & $3.4594$ & $3.4594$\\\hline
\end{tabular}
\end{center}
\label{coa1}
\end{table}%

\begin{table}[]
\caption{Number of times for divergence in $1000$ independent simulations.}
\begin{center}
\begin{tabular}{c||c|c|c|c|c|c|c|c|c}
\hline $(K,N)$ & \multicolumn{3}{c|}{$(3,9)$} & \multicolumn{3}{c|}{$(4,16)$} & \multicolumn{3}{c}{$(5,15)$}\\\hline
Equilibrium & NE & OSE & ASE & NE & OSE & ASE & NE & OSE & ASE\\\hline \hline
$x=5$ & $0$ & $0$ & $0$ & $5$ & $2$ & $0$ & $9$ & $7$ & $0$\\\hline
$x=0.8$ & $78$ & $58$ & $0$ & $310$ & $251$ & $0$ & $386$ & $323$ & $0$\\\hline
$x=0.5$ & $23$ & $25$ & $0$ & $115$ & $122$ & $0$ & $200$ & $203$ & $0$\\\hline
$x=0.2$ & $0$ & $0$ & $6$ & $0$ & $0$ & $39$ & $0$ & $0$ & $48$\\\hline
\end{tabular}
\end{center}
\label{coa2}
\end{table}

\begin{table}[]
\caption{Average users' sum-rates for the $3$-user $9$-subcarrier interference channel.}
\begin{center}
\begin{tabular}{c||c|c|c}
\hline Sum-rate & User $1$ & User $2$ & User $3$\\\hline\hline
NE & $13.9944$ & $13.8966$ & $13.8021$\\\hline
OSE & $16.5275$ & $15.5493$ & $15.4743$\\\hline
ASE & $32.6641$ & $32.8518$ & $33.0399$\\\hline
\end{tabular}
\end{center}
\label{coa3}
\end{table}%

\begin{table}[]
\caption{The average per-user sum-rates for ASE and ISB.}
\begin{center}
\begin{tabular}{c||c|c|c}
\hline
$(K,N)$ & $(3,9)$ & $(6,18)$ & $(12,36)$\\\hline\hline
ASE & $32.8414$ & $32.2201$ & $31.6874$\\\hline
ISB & $34.6674$ & $35.5500$ & $35.6335$\\\hline
\end{tabular}
\end{center}
\label{coa4}
\end{table}%

\end{document}